\documentclass[10pt]{amsart}
\usepackage{moreverb}
\usepackage[para]{threeparttable}
\usepackage{graphicx,amssymb,amstext,amsmath}
\usepackage{graphicx}
\usepackage{rotating}
\usepackage{natbib}

\def\s{\sigma^2}

\def\m{\mu}
\def\m1{\mu_1}
\def\m0{\mu_0}
\def\b{\beta}

\def\g{\gamma}

\def\bY{\mathbf Y}
\def\bV{\mathbf V}
\def\ij{_{ij}}

\def\eco{\mbox{eco}}
\def\epd{\mbox{epds}}
\def\eth{\mbox{eth}}
\def\age{\mbox{age}}
\title[Handling missing values in cost-effectiveness analyses of CRT]
{Handling missing values in cost-effectiveness analyses that use data from cluster randomised trials.}

\author[K. ~D\'iaz-Ordaz ]{Karla ~D\'iaz-Ordaz}
\address{Centre for Primary Care \& Public Health, Queen Mary University of London,
58 Turner Street, London E1 2AB, UK. }
\email{k.d.ordaz@qmul.ac.uk}

\author[M. ~Kenward]{ Michael G. ~Kenward}
\address{London School of Hygiene and Tropical Medicine,
Keppel Street, London, WC1E 7HT,
UK.}

\author[R.~Grieve]{Richard ~Grieve}
\address{London School of Hygiene and Tropical Medicine,
Keppel Street, London, WC1E 7HT,
UK.}

\date{} 
\keywords{missing data, multiple imputation, bivariate models, cost-effectiveness.\\{\it email for correspondence}: k.d.ordaz@qmul.ac.uk
}

\begin{document}

\begin{abstract}
Public policy-makers use cost-effectiveness analyses (CEA) to decide which health and social care interventions to
provide. Appropriate methods have not been developed for handling missing data in complex settings, such as for CEA
that use data from cluster randomised trials (CRTs). We present a multilevel multiple imputation (MI) approach that
recognises when missing data have a hierarchical structure, and is compatible with the bivariate multilevel models
used to report cost-effectiveness. We contrast the multilevel MI approach with single-level MI and complete case
analysis, in a CEA alongside a CRT. The paper highlights the importance of adopting a principled approach to
handling missing values in settings with complex data structures.
 \end{abstract}
\maketitle

\section{Introduction}

Public policy-makers use cost-effectiveness analyses (CEA) in deciding  which health and social care interventions
to prioritise \citep{NICE,CADTH2006,PBCA2008,IQWIG2009}. CEA exploit evidence from
randomised studies, and if they adopt appropriate statistical methods, can provide accurate assessments of which
interventions are most worthwhile \citep{Gold1996,OHagan2001c,Willan2006, Glick2007,Gray2010}. CEA raise major challenges for the analytical approach as the data tend to have complex structures, with correlated cost and effectiveness endpoints \citep{Willan2003,Willan2006a}, hierarchical data \citep{Manca2005b,Pinto2005}, and costs with right-skewed distributions \citep{Manning2006,Jones2011a}. Most CEA that use individual-level data have observations with incomplete information \citep{Noble2011}. Statistical methods have not been developed that can simultaneously address all these issues. Hence studies may fail to provide the unbiased, precise cost-effectiveness estimates that decision-makers require.

This paper is motivated by CEA that use data from cluster randomised trials (CRTs), but the approach we propose
addresses three issues of general relevance. The first issue is raised by the bivariate nature of the outcomes,
which implies the need for joint modelling. Here, one endpoint is highly skewed, but inferences about means are
still required on the original scales of measurement.  The second issue is that randomisation is at the cluster
level, which implies that the data are hierarchical. The third issue, and the focus of this paper, is the presence of missing data.

Approaches have been proposed for jointly modelling costs and health outcomes while acknowledging that individual
costs tend to have right-skewed distributions \citep{Thompson2005}. There is a large literature on methods for handling clustered data, see for example \cite{Hayes2009}, \cite{Eldridgebook}, \cite{Aerts2002}, \cite{Goldstein}. Methods for CEA alongside CRT include: a `two-stage' non-parametric bootstrap procedure \citep{Flynn2005b, Bachmann2007}; bivariate Generalised Estimating Equations with robust standard errors \citep{Gomes2011}, and bivariate multilevel models (MLMs). Amongst the MLMs proposed are bivariate Normal models estimated by maximum likelihood \citep{Gomes2011}, or with Bayesian Markov-Chain Monte Carlo (MCMC) methods \citep{Grieve2010, Bachmann2007}.

An outstanding issue is handling missing data in CEA with complex structures. For example, in a CRT, the prevalence
of missing endpoint data may differ according to individual and cluster-level characteristics (e.g cluster size).
CEA methods guidance recommends multiple imputation (MI) \citep{Blough2009, Briggs2003, Ramsey2005}, but most published CEA still use complete case analysis \citep{Noble2011}. While MI approaches have been proposed for handling missing data with a clustered structure \citep{CarpenterGold, Schafer}, no previous study has developed methods for handling missing hierarchical data in complex settings, such as those seen in CEA that use cluster trials.

 The aim of this paper is to develop and illustrate an overall approach to analysing studies which have: bivariate
 outcomes with one highly skewed endpoint, a clustered structure and missing data. We do this using MI within a
 frequentist paradigm. At the same time, we explore the implications of failing to acknowledge relevant features of
 the setup in the handling of the missing data, in particular the potential consequences of ignoring clustering in
 the imputation step, and departures from Normality. We also compare the results we obtain with those from an
 analysis restricted to those individuals with complete data.

  In Section \ref{sec:ponder}, we introduce our case study which is a typical CEA that uses CRT data. In Section \ref{sec:substantive}, we develop a simple modelling framework for a clustered bivariate pair of  outcomes, one of which has a potentially non-Normal distribution. Section \ref{sec:missing} considers in some detail the handling of missing data in this setup, and explores the use of multilevel MI for this problem. In  Section \ref{sec:results}, we compare the results obtained from a range of alternative strategies. We close with a  discussion in Section \ref{sec:discussion}.

\section{Motivating example: The PoNDER study}\label{sec:ponder}
The PoNDER study (psychological interventions for post-natal depression trial and economic
evaluation) was a CRT evaluating an intervention for preventing postnatal depression,   \citep{Morrell2009}. It included 2659 patients who attended 101 primary care providers in the UK (general practices). Clusters were
randomly allocated to provide either usual care (control) or an intervention delivered by a health visitor
(treatment).  The intervention comprised health visitor training to identify and manage patients with postnatal
depression. As is common, the PoNDER CRT had an unbalanced design; the number of patients per cluster varied widely (from 1 to 101 in the control group and from 1 to 81 in the treatment group).

Patients were followed up for 18 months with costs ($\pounds$ sterling) and health-related quality of life (HRQoL)
recorded at six monthly intervals. This paper considers costs and HRQoL reported at six months. These HRQoL data
were used to adjust life years and present quality-adjusted life years (QALYs) over six months. Intra-cluster
correlation coefficients (ICCs) were moderate for QALYs (ICCq=0.04), but high for costs (ICCc=0.17).  While QALYs
were approximately Normally distributed, costs were moderately skewed.

Baseline measurements were collected from mothers at six weeks post-natally, for variables anticipated to be
prognostic for either cost or effectiveness endpoints.

Table \ref{Tabmiss} reports the percentage of observations with missing data, by treatment group. For each baseline
variable, less than $2.5\%$ of participants had missing data, but a relatively high proportion of individuals had
missing data for the cost endpoint; 31 clusters were without any observed cost data (15 in the control arm,
including one cluster that withdrew from the study).
\begin{table}
\caption{\label{Tabmiss} Description of missing data in the PoNDER case study, by treatment group.}
\centering
\begin{tabular}{lllllll}\hline
&&&\multicolumn{2}{l}{\small{Control group}} & \multicolumn{2}{l}{\small{Intervention group}}\\
&&&\multicolumn{2}{l}{\small{(Total n=911)}} & \multicolumn{2}{l}{\small{(Total n=1730)}}\\
\hline
\small{\it Outcome variables}&type&symbol&\small{Missing n}&$\%$&\small{Missing n}& $\%$\\ \\
Cost&continuous  &$c\ij$ &402 &41.1&460&26.6\\ QALY&continuous &$q\ij$ &39&4.3&59&3.4\\ \\
\small{\it Baseline variables} &   && & & \\ \\
\small{Edinburgh Postnatal}\\ \small{Depression Scale}&continuous &$\epd_{ij}$ &0 &0&0&0\\
\small{Ethnicity}&binary & $\eth_{ij}$&0&0&0&0\\
\small{Economic status}&binary&$\eco_{ij}$ & 0 & 0&0&0\\
Age&continuous &$\mbox{age}_{ij}$ &   1&0.1&0&0\\
\small{English as first language}&binary&$\mbox{eng}\ij$& 0&0&0&0\\
\small{Living alone}&binary&$\mbox{al}\ij$ &9&1.0&7&0.4\\
\small{Partner's economic status}&ordinal&$\mbox{pe}\ij$ &    7& 0.8&10 &0.6\\
\small{Benefits}&binary& $\mbox{b}\ij$&19&2.1&38&2.2\\
\small{History of depression}&binary&$\mbox{d}\ij$&5&0.5&6&0.3\\
\small{Any major life events}&binary&$\mbox{l}\ij$&8&0.9&9&0.5\\
\small{Relationship with baby}&ordinal & $\mbox{rb}\ij$  &12&1.3&20&1.2\\
\hline
\end{tabular}
\end{table}

The CEA presents incremental QALYs and costs as the differences in means, between the treatment and control groups
(\citealt{MorrelHTA}). Cost-effectiveness is then reported as incremental net monetary benefits (or INB, see equation
\eqref{inbdef} for definition).

To simplify the exposition, we restrict our analyses to those individuals with positive costs, by excluding 18
observations with zero costs (15 in the treatment group). See Section \ref{sec:discussion} for further discussion.

\section{Substantive model}\label{sec:substantive}

Let $C_{ij}$ and $Q_{ij}$ be the cost and QALY outcomes respectively from the $j$th patient in cluster $i$ of a
two-armed CEA alongside a CRT. We assume that the observations from different clusters are independent.

We are principally concerned with estimating the linear additive effect of treatment on mean costs and health
outcomes, with no additional covariates. Because of the simplicity of our setup, we are able to model the data from
the two treatment groups entirely separately, and then make the comparison. So, in the following, we show the
development for one treatment group; exactly the same arguments apply to the other.

First, we introduce bivariate Normal latent variables $\{u_{i},w_{i}\}$ to represent possible cluster effects for
cost and QALYs respectively, with
\begin{equation}\label{clustereffectsrho}
\left(\begin{array}{c}
u_{i}\\ w_{i}\end{array}\right)
\sim
\mbox{N}\left[\left(\begin{array}{c} 0\\ 0\end{array}\right),
\left(\begin{array}{c c}\s_{u} & \rho\sigma_u\sigma_w \\ \rho\sigma_u\sigma_w & \s_{w} \end{array}\right)
\right],
\end{equation}
where $\s_u, \s_w,$ and $\rho$ are the variances and correlation of the two latent variables respectively.

We now build the bivariate substantive model on the expectations of the two outcomes, $C_{ij}$ and $Q_{ij}$, defined
conditionally on the two cluster effects, first for cost:
\begin{equation}\label{eq:cost}
\mu_C=\mbox{E}[C_{ij} \mid u_i, w_i] = \b_1+ u_i
\end{equation}
with $\b_1$ the mean appropriate for the first treatment group, and then for QALYs, conditional on the costs and
cluster effects
\begin{equation}\label{eq:qualy}
\mu_Q=\mbox{E}[Q_{ij}\mid c_{ij}, u_i, w_i]= \g_1+\alpha c_{ij}+w_i,
\end{equation}
with $\g_1$ the intercept for $Q_{ij}$ for the first treatment group, and $\alpha$ the corresponding regression
coefficient for the costs.

We now introduce distributions for $C_{ij}$ and $Q_{ij}$, conditional on the cluster effects. It is assumed that
the conditional distribution of  $Q_{ij}$ given $C_{ij}$ is Normal, with variance $\sigma^2_q$. We consider three
possible distributions for $C_{ij}$: Normal, Lognormal and Gamma. Other distributions could of course also be
considered if thought appropriate. The choice of the Normal is straightforward, the mean is given by
(\ref{eq:cost}), with some variance $\sigma_c^2$ say. The Gamma alternative is introduced with a parameterisation
that implies that the coefficient of variation, $\sqrt{\eta}$ say, is constant across clusters; in contrast to the
Normal which implies constant variance. For $\mu_C$, the conditional mean as given in (\ref{eq:cost}), the chosen
Gamma density can then be written
\begin{equation}\label{eq:gammapdf}
f_C(c) = \frac{1}{\Gamma(\eta)}\left(\frac{\eta}{\mu_C}\right)^{\eta}x^{\eta-1}\exp(-\eta x/\mu_C).
\end{equation}
To maintain comparability with the Gamma distribution, we introduce the Lognormal with a somewhat unusual
parameterisation, in which the coefficient of variation is again constant across clusters. This gives
\begin{equation}\label{eq:lgnormpdf}
f_C(c)=\frac{1}{c\sqrt{2\pi \log(1+\eta)}}\exp{\left\{- \frac{(\log c-\mu_C)^2}{2\log(1+\eta)} \right\}.}
\end{equation}
We assume that, conditional on the cluster effects, $(C_{ij},Q_{ij})$ is independent of $(C_{ij'},Q_{ij'})$  for $j\neq j'$, and so the required joint
density, still conditional on the cluster effects, can be obtained from the product of the densities for $[ C_{ij}
\mid u_i, w_i ]$ and $[Q_{ij} \mid C_{ij}, u_i, w_i]$.

 Finally, to obtain the marginal likelihood for the data for one treatment group, it is then necessary to
combine this joint density over all relevant patients, and then integrate over the distribution of the cluster
effects. This needs to be done numerically. There are several approaches for this, here we have used adaptive
Gaussian quadrature as implemented in SAS PROC NLMIXED. We provide sample code for this in Appendix
\ref{appendix:sas}.

Using conventional likelihood procedures we can then obtain estimated means for cost and QALYs ($\hat{\mu}_{C,k}$
and $\hat{\mu}_{Q,k}$ say) for treatment groups $k=1,2$  respectively, together with their estimated variances and
covariances. Note that the separate modelling steps for the two treatment groups implies that estimates are
independent between groups. The increments between the two groups are then estimated as $\hat{\delta}_C =
\hat{\mu}_{C,2} - \hat{\mu}_{C,1} $ and  $\hat{\delta}_Q = \hat{\mu}_{Q,2} - \hat{\mu}_{Q,1}$.

The relative cost-effectiveness of treatment 2 against treatment 1 can be summarised by the INB defined as
\begin{equation}\label{inbdef}
\mbox{INB}(\lambda) = \lambda \delta_Q-\delta_C
\end{equation}
for $\lambda$, a given threshold willingness to pay for a unit of health gain. Its standard error can be calculated from the estimated variances and covariances of $\hat{\delta}_C$ and $\hat{\delta}_Q$  in the usual way.

\section{Missing Data}\label{sec:missing}

\subsection{Handling the missing data}

It is well known that missing data can be the source of selection bias,
and we are rarely able to construct analyses in which we can be confident that such bias has
been eliminated. Rather, we use what information is available both in the data and the substantive
setting in an attempt to reduce potential bias. Following this, carefully targeted sensitivity
analysis can play an valuable role. There are many ways in which analyses can attempt to deal
with missing data, and in which sensitivity analysis can be constructed, see for example \cite{lr02} and \cite{MKbook}.

One important source of information that can be used to potentially reduce bias
is contained in observed variables that are associated both with the outcome and with the
missing value process itself. If these variables are not part of the substantive model, they
are termed {\it auxiliary} variables in the missing value context.

There are several potential auxiliary variables in the current setting, and we will use an approach which can incorporate them. To explain the intended role of these variables we need to introduce some definitions due
to \cite{rubin76}. We use these in a fairly loose way here, more formal expositions can be found
in the books mentioned above, and in the references given there. One important distinction here
from Rubin's original definitions, is our use of these terms in a frequentist framework, which
implies rather stronger conditions than Rubin's likelihood based definitions.

Let $\bY_{ij} = \{Y_{ij1},Y_{ij2}\} = \{C_{ij},Q_{ij}\}$ be the pair of observations from subject $(i,j)$
and define the random variable $R_{ijl}$ to take the value 1 if $Y_{ijl}$ is observed and 0 if missing.
We say that the missing data are {\it Missing Completely at Random} (MCAR) if
$R_{ijl}$ and $Y_{ijl}$ are independent. By contrast, the data are {\it Missing at Random} (MAR) if there are
observed variables, contained in  $\bV$ say, such that $R_{ijl}$ and $Y_{ijl}$ are conditionally independent given $\bV$. It can be seen that MCAR implies MAR. We can reject the MCAR assumption in favour of MAR if we see associations between observed variables and $R_{ijl}$, which is of course completely observed.

If neither MCAR nor MAR hold, we say that the missing data are {\it Missing Not at Random} (MNAR). It
usually impossible to rule out MNAR in practice from the data at hand, because this depends critically on the
existence of associations between {\em unobserved} variables and the $R_{ijl}$, which the observed data cannot
exclude. It is this dependence between $R_{ijl}$ and $Y_{ijl}$ that is the potential source of bias.

It is therefore usually sensible to try at least to reduce this dependence by identifying potential auxiliary variables
from among those observed, and this will form the first step in handling the missing data. This will be done separately
for the two outcomes, because it is entirely plausible that very different missing value mechanisms will operate
with the two outcomes. We make the simplifying assumption that our auxiliary variables are completely observed. This is
not strictly necessary, and in principle the approach used here can be extended to the situation when they are not,
but for our present purposes the restriction to complete variables permits a simpler exposition.

\subsection{Multiple Imputation}

Having identified potential auxiliary variables, it is necessary to incorporate them into the analysis. If these variables were part of the substantive model, we could simply include them and so condition on them, and in this way reduce or remove the unwanted dependence between $R_{ijl}$ and $Y_{ijl}$. But as auxiliary variables, they are not in the substantive model, so this route is not available to us. Alternatively, we could construct an overall joint model in which these auxiliary variables are included as additional outcome variables. In the current setting this is awkward, although not infeasible, because of the clustered structure.

We instead choose to use multiple imputation \citep{rubin78, Kenward2007}. This has the advantage of retaining the original substantive model, adding to this an {\em imputation model}. This is essentially determined by the conditional distribution of the missing data given the observed data, which we allow to differ between outcomes and treatment arms.

In the present setting, in which we are only considering  missing data in the outcome, the conditional model follows from the substantive model. We note the role of the clustering in this: the observations from one cluster are mutually dependent, and so the conditional distribution of a missing value involves all the other observed values in the same cluster. To this basic model (or models) we add the auxiliary variables, again acknowledging the clustered structure.

Given the substantive and imputation models, conventional MI procedures can be followed. These are set out in detail in many references, including \cite{lr02} and \cite{MKbook}. The overall MI procedure is as follows.\\[1ex]

\noindent
(1) The imputation model is fitted to the observed data and Bayesian draws are taken from the posterior
of the model parameters.\\[1ex]
\noindent
(2)
The missing data are imputed from the imputation model, using the parameters drawn in step (1).\\[1ex]
\noindent
(3)
The substantive model is fitted (here using maximum likelihood) to the data set that has been {\em completed} using the imputations from step (2),
producing parameter estimates and their estimated covariance matrix.\\[1ex]
\noindent
(4)
Steps (1)-(3) are repeated a fixed number, $K$ say, of times.\\[1ex]
\noindent
(5)
The $K$ sets of parameter and covariance estimates from step (3) are then combined using Rubin's formulae \citep{rubin87} to produce a single MI estimate of the substantive model parameters and associated covariance matrix.

Under the MAR assumption, this will produce consistent estimators and in the absence of auxiliary variables, is asymptotically (as $K$ increases) equivalent to maximum likelihood.

For the current analysis, the whole MI procedure has been done separately for the two treatment groups. To carry this
out, we do need the facilities to make the required Bayesian draws from the imputation model, which is bivariate and includes a cluster random effect.
One route for this is the MLwiN procedures developed by \cite{CarpenterGold}. This is restricted however to the multivariate Normal distribution, and the imputation step has been undertaken on the log of the costs, transforming back
to the original scale in step (3) above. The approximation implied  under the Lognormal and Gamma substantive models is unlikely to be critical. The more flexible multilevel imputation procedure of \cite{Carpenter2011} might be considered for future work, or a bespoke sampler developed for the specific bivariate models used here.

\subsection{Specifying the imputation model for the PoNDER case study}\label{PonderIM}
To investigate the associations of observed variables with the $R_{ijl}$, it is natural to use logistic regression, in this example with and without random cluster effects. This has been done separately here for cost ($l=1$) and QALYs ($l=2$) and also for each treatment group ($k=1,2$).

In addition to the patient-level covariates described, we added the cluster-level variable {\it cluster size, $n_i$},
defined as the number of participants randomised in each cluster. Previous studies suggest that cluster size may be associated with costs or health outcomes \citep{Campbell2000, Omar2000, Neuhaus2011}.  We also consider that the number of participants recruited in each cluster may be associated with missingness. In PoNDER, because clinical protocols were less restrictive in the control than treatment group, it was anticipated that any relationship between the cluster size and the endpoints would be stronger in the control group.

In the control group,  before allowing for clustering, ethnicity, economic status and cluster size were associated with missing costs, but after including a random effect to allow for clustering no covariates were associated with missing costs. Cluster size and $\epd$ seem to be associated with unobserved QALYs. In addition, $\epd$ and cluster size ($n_i$) are associated with costs and economic status, ethnicity and $\epd$ are associated with QALYs.
In the treatment group, cluster size was seen to be  associated with missing cost at the individual level,
while adjusting for clustering resulted in economic status being predictive of missing costs. Only age is predictive
of QALYs missingness, both ignoring and accounting for clustering. In addition, $\epd$ is associated with the value of both cost and QALYs.

The MI algorithm implemented here assumes all variables included in the model are multivariate Normally distributed. We exploit this by choosing the same imputation models for both outcomes, adding all auxiliary variables which seem associated with either endpoint and their missingness, and modelling the two outcomes simultaneously. The imputation models chosen are summarised in Table \ref{impmodels}.

\begin{table}
\caption{\label{impmodels} Single-level (ignoring clustering) and multilevel (accounting for clustering) imputation models used for the cost and QALY endpoints in the PoNDER case study. Models that included a cluster-level auxiliary variable are indicated by -C.}
\centering
\begin{tabular}{cccc}
\hline
Type& Model & Control Group & Intervention Group\\ \hline
Single level &SL& $\epd_{ij}+\eco_{ij}+\eth_{ij}$&$\epd_{ij}+\eco_{ij}+\age_{ij}$\\
& SL-C & $\epd_{ij}+\eco_{ij}+\eth_{ij}+n_i$&$\epd_{ij}+\eco_{ij}+\age_{ij}+n_i$\\
Multilevel & ML & $\epd_{ij}+\eco_{ij}+\eth_{ij}$ &$\epd_{ij}+\eco_{ij}+\age_{ij}$\\
&ML-C & $\epd_{ij}+\eco_{ij}+\eth_{ij}+n_i$ &$\epd_{ij}+\eco_{ij}+\age_{ij}+n_i$\\ \hline
\end{tabular}
\end{table}

\section{Multiple imputation estimates for the example data set}\label{sec:results}
The imputation models that ignore clustering (SL, SL-C) were implemented with the {\bf ice} command in STATA (by chained equations), while the multilevel imputation models (ML, ML-C) used multivariate Normal MCMC algorithms implemented in MLwiN {\bf mi} macros.

\begin{figure}[h]
\centering
\includegraphics[scale=0.7]{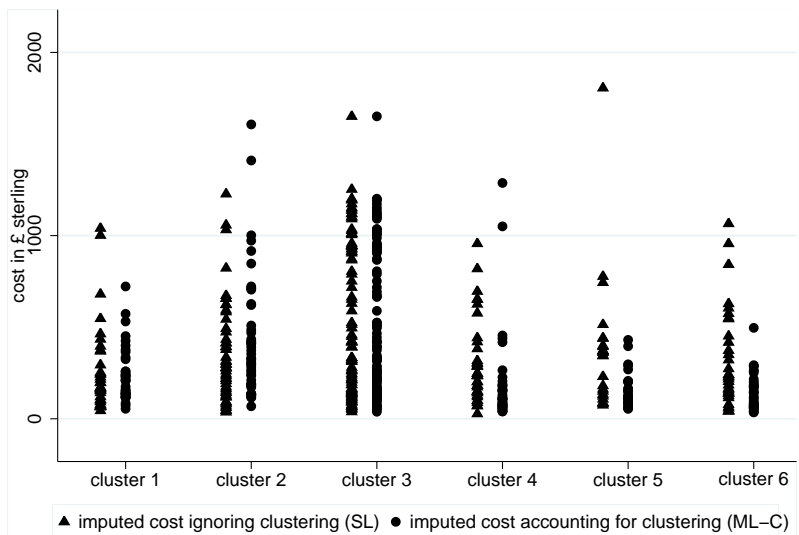}
\caption{\label{Fig2}Difference in the ``spread'' of imputed data, depending on whether the imputation model
ignored
or accounted for clustering (Model SL versus Model ML-C). The Figure shows the distribution of costs in the six
control clusters with the highest number of missing values in the original dataset.}
\end{figure}

For each imputation model in Table \ref{impmodels}, we obtained five imputed datasets. Figure \ref{Fig2} highlights
the impact that accounting for clustering in the MI model can have on the distributions of ``imputed'' values. It shows imputed cost data for the six clusters in the control arm with the highest number of observations
with missing cost data. The Figure contrasts data imputed after applying the single-level imputation models versus
the multilevel imputation which included cluster size as an auxiliary variable. The cost distribution appears
somewhat less clustered after the single-level imputation than after multilevel imputation.

The five multiply imputed datasets were each analysed with the three substantive models defined in Section
\ref{sec:substantive}, i.e. random cluster effects models with bivariate Normal (N-N), Lognormal-Normal (L-N) and
Gamma-Normal (G-N) distributions. Table \ref{miest} reports the MI estimates for mean cost and QALYs by treatment
arm, and for comparison also includes estimates from complete cases (CC).

Table \ref{miest} shows that, as anticipated, the standard errors for both endpoints are larger for the control than for the treatment group. It is also clear that ignoring the hierarchical structure of the data in the imputation model results in different point estimates for mean cost, especially in the control arm. For all approaches, the estimated correlations between cost and QALYs are small.

We use the estimates from Table \ref{miest} to obtain incremental costs, QALYs and INBs for a willingness to pay,  $\lambda$, of $\pounds 20000$ per QALY. These are reported in Table \ref{MItable}.

\begin{table}
\caption{\label{miest}Mean (SE) costs (in $\pounds$ sterling) and QALYs, and estimated correlations between the two endpoints. Estimates are according to choice of approach for handling missing data, and for alternative bivariate substantive models.}
\centering
\begin{tabular}{l lrrrrrr}
\hline
\multicolumn{2}{c}{Missing data approach}&\multicolumn{3}{c}{Control Group} & \multicolumn{3}{c}{Intervention Group}\\ \hline
Model&\small{Estimates} & N-N&L-N& G-N& N-N&L-N& G-N\\ \hline\hline CC&\small{Mean cost } & $273.3$  & $286.6$
&$277.2$    &$256.8$   & $258.5$ & $254.6$ \\ & (SE)& ($25.3$) & ($23.0$) &($24.8$)&($11.7$)&($11.6$)&($10.7$)\\
       & \small{Mean QALYs}& $0.027$ &$0.027$ &$0.027$& $0.030$&$0.030$&$0.030$ \\
& (SE)&(0.0014)&(0.0014)&(0.0014) &(0.0008)& (0.0008)& (0.0008) \\
         & \small{corr(c,q)}& $-0.03$&$ -0.03$ & $-0.04$ & $-0.04$&$-0.04$&$-0.04$ \\ \hline
SL &\small{Mean cost }&295.0& 299.1& 295.00 &251.4& 253.1&251.3 \\ & (SE)&(16.8) &(16.9)&(19.1)&(9.9)& (9.5)&
(9.1)\\
       & \small{Mean QALYs}&$0.027$ &$0.027$ &$0.027$& $0.030$&$0.030$&$0.030$ \\
& (SE)& ($0.0011$)&($0.0011$)&($0.0011$)&($0.0007$)&($0.0007$)&($0.0007$)\\
         & \small{corr(c,q)}&$0.00$&$-0.09$& $-0.06$& $-0.09$&$ -0.08$&$-0.06$ \\ \hline
SL-C& Mean cost & $268.1$&$275.5$& $270.4$ &$257.2$&$257.5$&$255.4$\\ & (SE)&
($18.8$)&($16.9$)&($18.1$)&($9.1$)&($9.2$)& ($8.8$)\\
       & Mean QALYs & $0.026$ &$0.026$ &$0.026$& $0.030$&$0.030$&$0.030$\\
& (SE)& ($0.0011$)&($0.0011$)&($0.0011$)&($0.0007$)&($0.0007$)&($0.0007$)\\
       & corr(c,q)&$0.01$&$0.05$&0.001&$-0.06$&$-0.08$&$-0.04$\\ \hline
ML& Mean cost &$264.6$& $265.0$&$262.5$&$256.9$& $257.9$&$255.1$\\ & (SE)&
($31.6$)&($25.8$)&($29.4$)&($12.8$)&($12.9$)& ($12.1$)\\
      & Mean QALYs & $0.026$ &$0.026$ &$0.026$& $0.030$&$0.030$&$0.030$\\
& (SE)& ($0.0011$)&($0.0011$)&($0.0011$)&($0.0007$)&($0.0007$)&($0.0007$)\\
       & corr(c,q)&$0.06$&$0.03$&$0.04$&$-0.07$&$ -0.07$& $-0.06$ \\ \hline
ML-C &Mean cost & $270.0$& $280.6$& $275.0$ &$262.1$& $262.5$&$259.5$\\ & (SE)&
($21.0$)&($23.4$)&($23.5$)&($14.0$)&($12.2$)& ($12.1$)\\
      &  Mean QALYs & $0.026$ &$0.026$ &$0.026$& $0.030$&$0.030$&$0.030$\\
& (SE)& ($0.0011$)&($0.0011$)&($0.0011$)&($0.0007$)&($0.0007$)&($0.0007$)\\
      & corr(c,q)& $-0.09$&$ -0.09$& $-0.08$& 0.01&$0.00$ & 0.04\\
\hline
\end{tabular}
\end{table}

\begin{table}
\caption{\label{MItable}Estimated incremental cost (in $\pounds$ sterling), QALYS and INBs at a threshold of $\pounds 20000$
per QALY. Estimates are according to choice of approach for handling missing data for alternative bivariate
substantive models}
\centering
\begin{tabular}{llrrr}
\hline
Estimates (SE)&Model & N-N  &  L-N & G-N  \\
\hline
Incremental cost & CC  &  $-16.5$ ($27.9$)  &  $-28.1$ ($25.8$)  &$-22.5$ ($27.00$)\\ 	$\delta c$&SL& $-43.6$
($19.5$) &$-46.0$ ($19.4$)  &$-43.7$ ($21.2$)\\ 	&SL-C &  $-10.9$ ($20.9$)  &$-18.0$ ($19.2$)  &$-15.0$
($20.1$)\\ 	&ML& $-7.7$ ($34.1$)  &$-7.1$ ($28.8$)  &$-7.4$ ($31.8$) \\ 	&ML-C & $-7.9$ ($25.2$)    &$-18.1$
($26.4$) &$-15.5$  ($26.4$)\\ Incremental QALYs  & CC& $0.003$ ($0.002$)& $0.003$ ($0.002$)&$0.003$ ($0.002$) \\
$\delta q$	&SL & $0.004$ ($0.001$) & $0.004$ ($0.001$) &$0.004$ ($0.001$)\\ 	&SL-C & $0.004$ ($0.001$) & $0.004$
($0.001$) &$0.004$ ($0.001$)\\ 	&ML & $0.004$ ($0.001$) & $0.004$ ($0.001$) &$0.004$ ($0.001$)\\ 	&ML-C & $0.004$
($0.001$) & $0.004$ ($0.001$) &$0.004$ ($0.001$)\\ INB &CC&$76.5$ ($43.7$) & $81.3$ ($41.6$) &$75.3$ ($43.5$)\\
 	&SL &$117.0$ ($34.0$)&  $117.4$ ($34.9$) &$117.9$ ($34.0$)\\
	&SL-C& $82.6$ ($33.8$)  &$94.0$ ($32.4$)  &$90.8$ ($33.3$)\\ 	&ML&$82.7$ ($42.8$)  & $82.5$ ($38.6$) &$82.6$
($41.2$)  \\ 	&ML-C& $84.5$ ($38.0$) &$96.0$ ($38.2$) &$93.0$ ($38.5$) \\
\hline
\end{tabular}
\end{table}

Table \ref{MItable} shows that the estimates of incremental cost, incremental QALYs and INB are relatively insensitive to the choice of cost distribution. In fact, for the incremental QALYs, where there is little missing data and the ICCs are low, the estimates and their standard errors are virtually identical following each missing data approach.

However, inferences about the estimated  incremental costs and the INBs differ depending on the approach for handling missing data.
Firstly, complete case estimates are likely to be biased, as the missing mechanism is probably not MCAR and the substantive model is not adjusting for any covariates. Single-level MI approaches produce smaller standard errors than those obtained with multilevel MI and CC. This is because cost has a large ICC and we are looking at a between-cluster estimator. As a consequence, there is an increased risk of type I error, regardless of the choice of cost distribution used for the substantive model.

Moreover, ignoring informative cluster size in the multilevel imputation model increases the magnitude of the estimated standard errors. This cluster-level covariate (cluster size) is associated with cost and with cost missingness, and so excluding it from the imputation model reduces the precision of the estimate, as information is lost. By contrast, including cluster size in  the single-level imputation model results in point estimates for the incremental cost which are similar to those following multilevel MI, although estimates for the standard errors are still smaller than the corresponding multilevel MI estimates.

\begin{figure}
\centering
\includegraphics[scale=0.5]{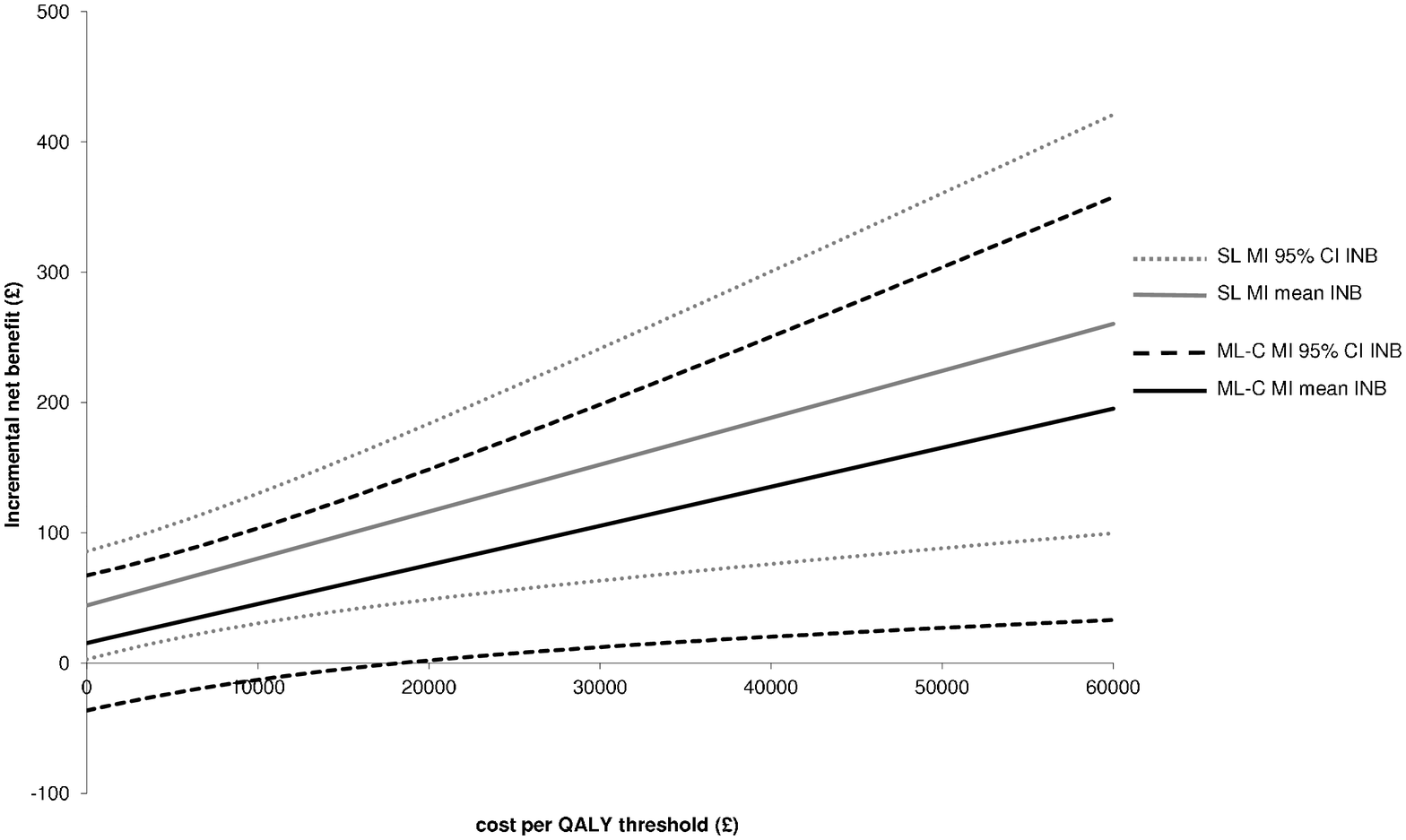}
\caption{\label{Fig1GN} MI Estimates of mean ($95\%$) incremental net benefit (INB) from model ML-C (multilevel
multiple imputation with informative cluster size) versus model SL (ignoring clustering) using bivariate Gamma-Normal substantive model}
\end{figure}

Figure \ref{Fig1GN} shows the INB (with $95\%$ CI) at alternative thresholds of willingness to pay for a QALY
gained. With the single-level MI, the INB and $95\%$ CI are positive throughout, indicating that the treatment is
cost-effective. For the multilevel MI that acknowledges informative cluster size, the $95\%$ CIs around the INB are
wider and include zero at realistic thresholds for a QALY gained. While for both approaches, the INB remains
positive throughout, the single level MI approach appears to overstate both the absolute level of the INB, and the
precision surrounding the estimate.

Hence, in the PoNDER case study, once a more appropriate approach is taken to handling the missing data, it is less
certain that the intervention is cost-effective.

 \section{Discussion}\label{sec:discussion}
This paper provides a principled approach to handling missing data with complex structures, exemplified by CEA
that use data from CRTs. The proposed approach follows the general principle that the imputation model should reflect the  structure of the analytical model. In the context of cluster trials, just as the substantive model can account for clustering with a MLM, so must the imputation model.
Moreover, because the analytical models typically used in CEA estimate the linear additive effects of treatment on mean costs and health outcomes without covariate adjustment, MI has particular appeal in this setting. By separating the imputation and substantive models, information on those auxiliary variables, such as baseline patient characteristics, associated with missingness and the endpoints of interest can and should be used, without the analyst having to modify the substantive model.

Our study highlights that a single-level imputation model can underestimate the uncertainty surrounding the estimates of interest. More generally,  \cite{Taljaard2008} showed that MI approaches that ignore clustering can increase Type I
errors. Another common approach to handling missing data in CRTs is to include cluster as a fixed effect in a single-level imputation model \citep{White2011, Graham2009}, but this does not produce an imputation model that properly captures the conditional distribution of the missing data given the observed. Indeed, including cluster as a fixed effect represents the limiting case where the ICC tends to one, and does not reflect the variability of the imputed values. The simulation study by \cite{Andridge2011} found that including cluster as a fixed effect in the imputation model, can overestimate the variance of the estimates, especially when ICCs are low, and there are few clusters. Both features are common in our setting, a recent review found that out of 63 published CEA alongside cluster trials, $40\%$ had fewer than 15 clusters per treatment arm, and one third reported ICCs of 0.01 or less for health outcomes \citep{Gomes2012}.

The case study presented here suggests that if there are cluster-level covariates
associated with the missingness patterns and the value of the outcomes to be imputed, then including them even in a single-level imputation model may potentially provide more accurate point estimates. This is because including
covariates that predict dependency on cluster reduces the ICC. However, unless such covariates fully explain the
between-cluster variance, such single-level MI approaches would still overstate precision. Hence, we propose
imputation models with random effects for clusters.

A general challenge in CEA is choosing appropriate statistical models for costs, which tend to have right-skewed distributions. The bivariate models developed here use marginal log-likelihoods for one outcome and
conditional for the other, by expressing the relationship between the two responses as a linear regression (see sample code provided in Appendix \ref{appendix:sas}). In principle, these models are generalisable to allow mixed
distribution log-likelihoods, provided the conditional likelihood of the dependent outcome is known explicitly and
can be optimised. The advantage of this approach is that, by parameterising the density according to the coefficient
of variation, and maximising the log-likelihood obtained, we avoid log-transforming and re-transforming costs in the
presence of heteroscedasticity  \citep{Manning1998, Duan1983, Mullahy1998, Manning2001}. We
consider three cost models that make alternative distributional assumptions, but keep an essentially Lognormal
imputation model throughout, and use standard optimisation routines to obtain maximum likelihood estimates. Our
findings suggest that assuming a different distribution for the imputation versus analytical model appears to have
little impact, whereas the choice of whether or not the imputation model accounts for clustering can be important.

A previous barrier to adopting principled MI approaches for hierarchical data was the lack of available software, but this is no longer the case. There are now three options for performing multilevel MI based on multivariate Normal MCMC algorithms:  PAN \citep{Schafer} which is available as an
{\bf R} package \citep{R}, the {\bf mi} macro \citep{CarpenterGold} which operates within MLwiN \citep{mlwin} and
can handle up to four hierarchical levels and binary variables, and REALCOM-impute macros \citep{Carpenter2011}, which  can also handle categorical variables and cluster-level variables with missing data.

The approach presented in this paper has some limitations. For simplicity, we assume the missing data mechanism is MAR throughout. However, MI provides a flexible and convenient route for investigating sensitivity to alternative MNAR mechanisms \citep[e.g.][]{Carpenter2007}, and in principle standard procedures should apply without much modification.
A further advantage of MI, which this case study could not exploit, is that the
imputation model may include post-randomisation variables associated with missingness and endpoints, which should not be
included in the substantive model.

 A further concern is that the imputation and analytical models may make incorrect distributional assumptions.
Simulation studies by \cite{SchaferBook} have shown that MI can be fairly robust to  model misspecification, but
their simulation settings did not include multilevel structures; \cite{Yucel2010} recently investigated the impact
of misspecifing the multilevel imputation model but focused on violations of the distributional assumptions for
 the random-effects. They find that when the imputation model has sufficient auxiliary variables, inferences are
 insensitive to non-Normal random-effects, unless the rates of missingness are very high or the sample size is
 small. They obtained similar results when the assumption that level-1 residuals were Normally distributed was
 violated.

While we propose a general approach to handling missing data in cluster trials, it is illustrated through a single
case study which cannot represent all the circumstances faced by CEA that use CRTs. There may be circumstances when
the data display quite different structures to those considered here, for example where there are a high proportion with
zero costs \citep{Mullahy1998}, QALYs with highly irregular distributions \citep{Basu2012}, or there are many
auxiliary variables available.

This paper suggests further extensions. Here, we combine multilevel MI with a MLM estimated by maximum likelihood, but there may be circumstances where it would be advantageous to combine multilevel MI with MLM estimated by Bayesian MCMC \citep{Lambert2005}, for example when synthesising evidence across
multiple sources \citep{Welton2008}; or indeed adopt a fully Bayesian approach to handling the missingness and specifying the analytical models \citep{Mason2012}.

Future simulation studies could be useful in contrasting the relative performance of the alternative approach across
a broad range of settings including those where there are a high proportion of observations with zero costs,  health
outcomes with irregular distributions, and few clusters. Clearly, ignoring clustering in the imputation model will
have less impact as the ICC decreases. One way of reducing the outstanding variation at the cluster-level within a
single-level imputation model is to introduce more cluster-level covariates. Further work is needed to assess under
what circumstances this simple MI approach would provide reliable inferences.

Finally, it would also be useful to extend the approaches to handling missing data to other settings with
hierarchical data. These could include trials with repeated measures over time, studies with a high proportion of
zero costs, censored costs, or non-randomised studies where covariate adjustment is required.

\section*{Acknowledgments}

We are grateful to  Jane Morrell (PI) and Simon Dixon for permission to use, and for providing access to, the PoNDER
data. We thank James Carpenter, Simon Thompson, Richard Nixon, John Cairns, Manuel Gomes and Edmond Ng for helpful
discussions. KDO was supported by a NIHR Research Methods Fellowship, and RG was partly funded by the UK Medical
Research Council.
\appendix
\section{Implementation in SAS}\label{appendix:sas}
 We have developed a method that allows us to exploit the optimisation of general likelihood functions available in
 SAS procedure NLMIXED. Briefly, we duplicated the data and created an indicator variable for the first copy of the
 data, $\mbox{flag}=1$. We then used an $\mbox{if}$ statement indicating we wished to estimate cost parameters if
 $\mbox{flag}=1$.

With this method, we were able to use marginal expressions of corresponding log-likelihood to estimate parameters
for costs using in turn either a Normal, Lognormal or Gamma log-likelihood; the last two parameterised by the
coefficient of variation. We use Gauss-Hermite quadrature, with 70 quadrature points, and the  Newton-Raphson
maximisation technique to estimate the maximum-likelihood parameters. As likelihood maximisation is sensitive to the
initial parameters chosen in the NLMIXED model, we ran this twice, using different initial values, to ensure
optimization had achieved convergence.

Sample SAS code can be found below.
\begin{verbatim}
proc nlmixed data=ponder2 method=Gauss  qpoints=70 cov corr;
title "Control group bivariate Gamma-Normal with 2 Cluster Effects";
where group=0;
x=cost;
y=qalygain;
parms  b0=268 c0=0.27  a=3 lsyx=-7 lnsc=8 lnse=2 r=0.01;
mux= b0+u1; varyx = exp(lsyx);
muyx= c0+u2+(varx*(a/mux**2))*x;
if (flag=1) then
ll=-x*a/mux+(a-1)*log(x)-a*log(mux)+a*log(a)-log(Gamma(a));
else ll=-(1/2)*log(2*constant('pi'))-log(varyx)-((y-muyx)**2)/(2*varyx);
if (flag=1) then z = x;
else z = y;
 model z ~ general(ll);
random u1 u2 ~ normal([0,0],[exp(lnsc),r, exp(lnse)]) subject=cluster;
estimate "my" c0+(varyx*(a/b0**2))*b0; run;
\end{verbatim}



\end{document}